\documentclass[12pt, a4paper]{article}

\usepackage{graphicx}
\usepackage{amssymb}
\usepackage{amsmath}
\usepackage{bm}
\usepackage{color}
\usepackage{theorem}
\usepackage{subcaption}

\usepackage[sort&compress,numbers, merge]{natbib}

\definecolor{Orange}{cmyk}{0,0.61,0.87,0}
\definecolor{JungleGreen}{cmyk}{0.99,0,0.52,0}
\definecolor{OliveGreen}{cmyk}{0.64,0,0.95,0.40}
\definecolor{Brown}{cmyk}{0,0.81,1,0.60}
\definecolor{RoyalBlue}{cmyk}{0.71,0.53,0,0.12}

\usepackage[colorlinks=true, linkcolor=black, citecolor=black,
urlcolor=black]{hyperref}

\begin{document}

\begin{titlepage}

\begin{flushright}
{\tt 
UT-17-07\\
KYUSHU-RCAPP 2017-02 
}
\end{flushright}

\vskip 1.35cm
\begin{center}

{\Large
{\bf
 Extending the LHC Reach for New Physics  \\[5pt]
with Sub-Millimeter Displaced Vertices
}
}

\vskip 1.2cm

Hayato Ito$^a$,
Osamu Jinnouchi$^b$,
Takeo Moroi$^{a}$,
Natsumi Nagata$^a$,
and
Hidetoshi Otono$^c$

\vskip 0.8cm

{\it $^a$Department of Physics, University of Tokyo, Tokyo
 113-0033, Japan}\\[3pt]
{\it $^b$Department of Physics, Tokyo Institute of Technology, Tokyo 152-8551,
 Japan}\\[3pt]
{\it $^c$Research Center for Advanced Particle Physics, Kyushu University,\\
 Fukuoka 819-0395, Japan}

\date{\today}

\vskip 1.5cm

\begin{abstract}

  Particles with a sub-millimeter decay length appear in many
  models of physics beyond the Standard Model.
  However, their longevity has
  been often ignored in their LHC searches and they have been regarded
  as promptly-decaying particles.  In this letter, we show that, by
  requiring displaced vertices on top of the event selection criteria
  used in the ordinary search strategies for promptly-decaying
  particles, we can considerably extend the LHC reach for particles
  with a decay length of $\gtrsim 100~\mu{\rm m}$.  We discuss a way
  of reconstructing sub-millimeter displaced vertices by exploiting
  the same technique used for the primary vertex reconstruction on the
  assumption that the metastable particles are always pair-produced
  and their decay products contain high-$p_{\rm T}$ jets. 
  We show that, by applying a cut based on displaced vertices 
  on top of  
 standard kinematical
  cuts for the search of new particles, the LHC reach can be 
  significantly extended if the decay length is $\gtrsim 100~\mu{\rm m}$.
  In addition, we may measure the lifetime of the target
  particle through the reconstruction of displaced vertices, which
  plays an important role in understanding the new physics behind the
  metastable particles.

\end{abstract}

\end{center}

\end{titlepage}

\renewcommand{\thefootnote}{\#\arabic{footnote}}
\setcounter{footnote}{0}

\section{Introduction}
\label{sec:intro}

New metastable massive particles are predicted in a variety of
extensions of the Standard Model (SM) \cite{Fairbairn:2006gg}, and
have been explored at colliders such as the LHC. If these particles
have a decay length ({\it i.e.}, the product of the lifetime $\tau$
and the speed of light $c$) of ${\cal O}(1)$~m or shorter, then their
decay can occur within tracking detectors and thus it is in principle
possible to directly observe their decay points, which are away from
the production point. In fact, such attempts have been made in the LHC
experiments. For example, the ATLAS collaboration has searched for
displaced vertices (DVs) that originate from decay of long-lived
particles by investigating charged tracks with a transverse impact
parameter, $d_0$, of $2~{\rm mm} < |d_0| < 300$~mm, requiring that the
transverse distance between DVs and any of the primary vertices be
longer than $4~{\rm mm}$ \cite{Aad:2015rba}. This search is therefore
sensitive to metastable particles whose decay length is $c\tau \sim
\mathcal{O}(1-1000)~{\rm mm}$.  The disappearing-track searches
\cite{Aad:2013yna} can also probe a long-lived charged particle when
it decays into a neutral particle which is degenerate with the charged
particle in mass \cite{Feng:1999fu, Ibe:2006de, Asai:2007sw,
  *Asai:2008sk, *Asai:2008im, Nagata:2017gci}; the target of these
searches is $c\tau \gtrsim 10$~cm.

On the contrary, particles with a sub-millimeter decay length have
been beyond the reach of these searches. Such rather short-lived
particles have been often regarded as promptly-decaying particles and
probed without relying on their longevity.  Exceptionally, recently,
Ref.\ \cite{Khachatryan:2016unx} considered $R$-parity violating
supersymmetric (SUSY) model to which ``ordinary'' search strategies
does not apply, and showed that DV-based cuts may be useful for the
LHC study of such a model if the decay length of the lightest SUSY
particle is longer than $\mathcal{O}(100)\ \mu{\rm m}$.  From the
point of view of physics beyond the SM, however, there are a 
variety of well-motivated new particles with $c\tau\sim$ sub-millimeter besides 
the above case;
although LHC constraints on some of those have been already stringent even with
the analysis assuming that they decay promptly, 
inclusion of DV-based cuts upon it significantly extends 
the reach of those.
One of the
important examples for such particles is metastable gluino in SUSY models with heavy
squarks \cite{Wells:2003tf, *ArkaniHamed:2004fb, *Giudice:2004tc,
  *ArkaniHamed:2004yi, *Wells:2004di, Hall:2011jd, *Hall:2012zp,
  Ibe:2006de, 
  *Ibe:2011aa, *Ibe:2012hu, *Arvanitaki:2012ps, *ArkaniHamed:2012gw,
  *Evans:2013lpa, *Evans:2013dza}. In particular, if the squark 
masses are as heavy as the PeV scale, the decay length of the gluino
can be $c\tau\sim {\cal O}(100)\ \mu {\rm m}$ (assuming that the
gluino mass is around TeV) \cite{Toharia:2005gm, *Gambino:2005eh,
  *Sato:2012xf}. Metastable SUSY particles are also found in the
gauge-mediation models \cite{Giudice:1998bp, Draper:2011aa,
  *Evans:2016zau}, where the decay length of the next-to-lightest SUSY
particle decaying into gravitino can be order sub-millimeter, as
well as in $R$-parity violating SUSY models \cite{Barbier:2004ez,
  Graham:2012th, *Barry:2013nva, *Csaki:2013jza, *Csaki:2015uza}. In
addition, theories of Neutral Naturalness \cite{Chacko:2005pe,
*Burdman:2006tz, *Cai:2008au, Chacko:2015fbc},
hidden-valley models \cite{Strassler:2006im}, composite Higgs models
\cite{Barnard:2015rba}, dark matter models \cite{Chang:2009sv,
  *Co:2015pka}, and models with sterile neutrinos \cite{Basso:2008iv,
  *Helo:2013esa, *Izaguirre:2015pga} predict metastable particles with
an ${\cal O}(100)~\mu{\rm m}$ decay length.

In this letter, we discuss a method of searching for metastable
particles with DVs that is sensitive to $c\tau \lesssim 1$~mm as
well. Here, we focus on the cases where the target metastable
particles are always pair-produced, which is assured if the new
particles have a conserved quantum number; for instance, in $R$-parity
conserving SUSY models, SUSY particles, being $R$-parity odd, are
always produced in pairs.  In these cases, there are two decay points
in each event, which are separated from each other by order of the
decay length of these particles. We reconstruct these decay vertices
in a similar manner that is used for the primary vertex
reconstruction.  As shown below, using this method, we can distinguish
the decay vertices if these two are separated by $\gtrsim 100~\mu{\rm
  m}$. This method therefore enables us to search for sub-millimeter
DVs. By requiring DVs in addition to the event selection criteria used
in the promptly-decaying particle searches, we can go beyond the reach
of these searches if the target particle has a decay length of $c\tau
\gtrsim 100~\mu{\rm m}$. Moreover, when they are discovered, 
it is also possible to measure the
typical distance of the two decay points and thus to estimate the
decay length of the particles, which can provide important information
about the nature of the new physics behind the new particles.  To be
specific, in this letter, we consider metastable gluinos as an example
and demonstrate that the study of sub-millimeter DVs can significantly
enlarge the parameter region covered by the LHC.

The organization of this letter is as follows.  We first summarize how
the vertex is reconstructed in Sec.\ \ref{sec:vtx}.  In Sec.\
\ref{sec:gluino}, we describe our method to search for DVs.
  Then, we
apply our method to the metastable gluino search and show that the LHC
reach can be extended with the study of the DVs.  We also point out 
that it is possible to measure the decay length of the metastable particle by means of 
DV reconstruction. 
Sec.~\ref{sec:concl}
is devoted to conclusions and discussion.

\section{Vertex Reconstruction}
\label{sec:vtx}

First, let us briefly summarize how vertices are reconstructed at the
LHC experiment.  In order to make the argument clear, we use the
performance of the ATLAS detector. 
In this letter, we
concentrate on the case where a number of charged particles are
emitted from vertices, which is the case when the production and the
decay of metastable colored particles, like gluino, occur.  Then, with
the precise tracking of the charged particles by inner tracking
detectors, the decay vertex of the parent particle may be
reconstructed.

\begin{figure}[t]
  \centering
  \subcaptionbox{\label{fig:resolution_x} $x$-direction}{
  \includegraphics[width=0.48\columnwidth]{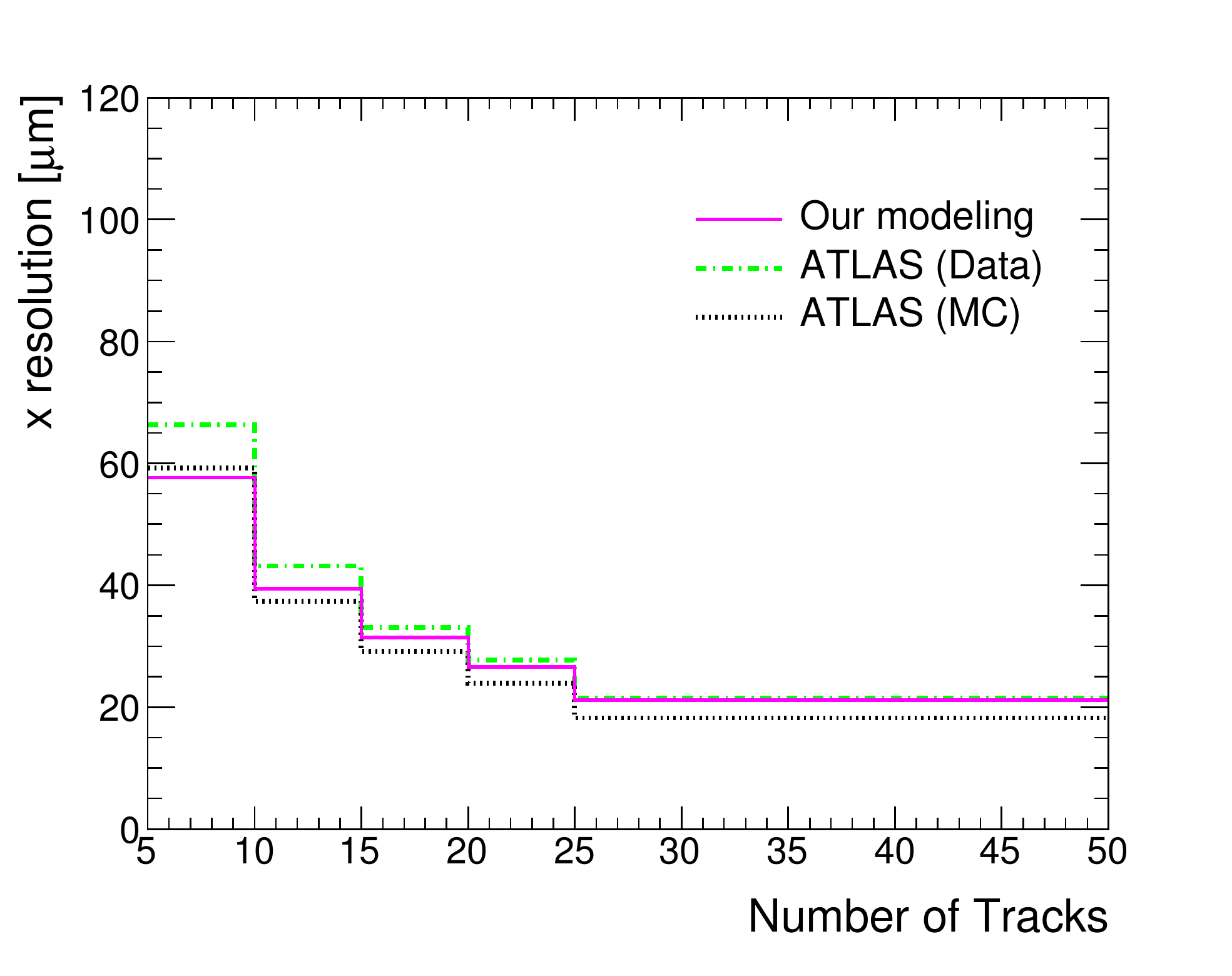}}
  \subcaptionbox{\label{fig:resolution_z} $z$-direction}{
  \includegraphics[width=0.48\columnwidth]{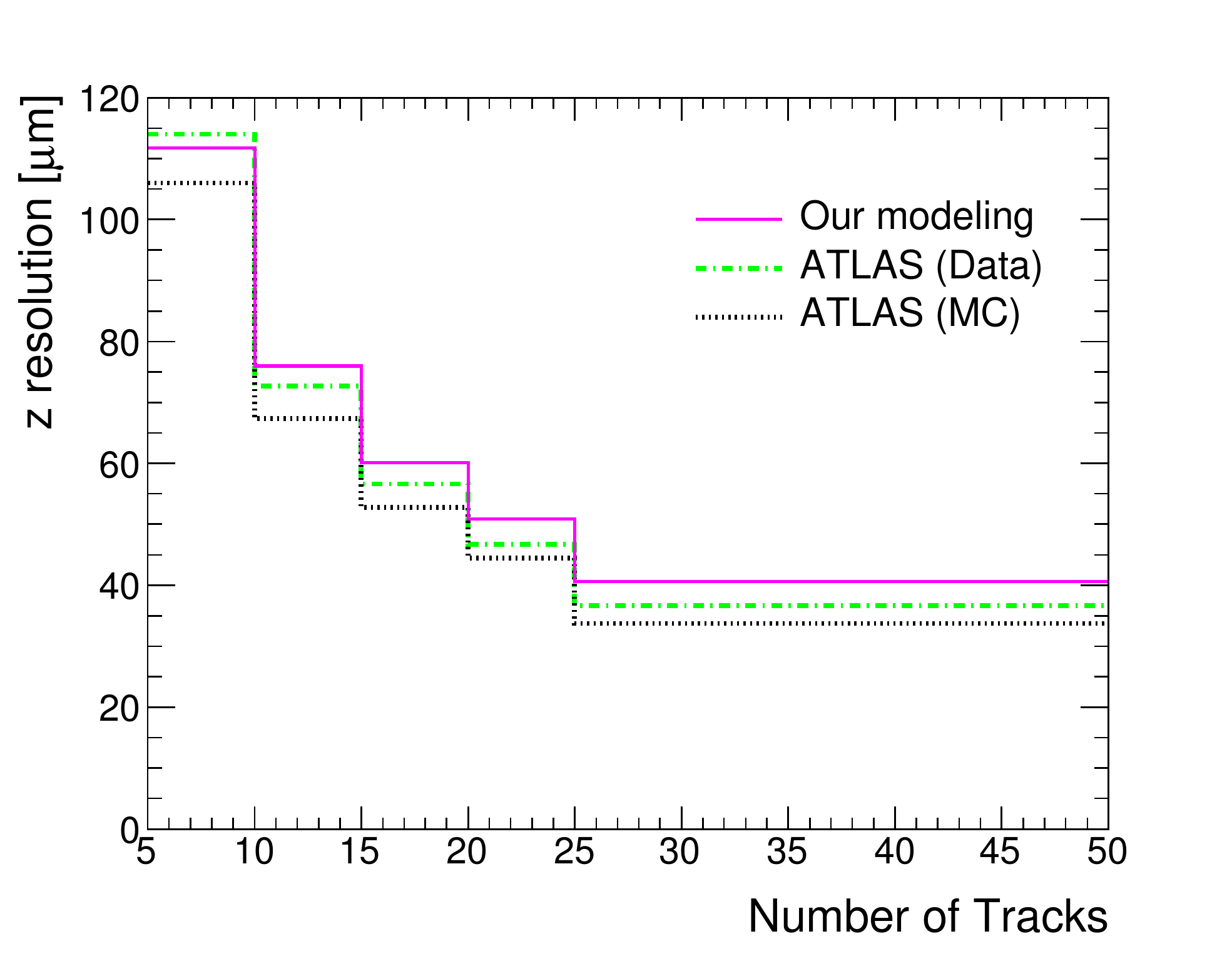}}
\caption{\small The resolutions of reconstructed primary vertex
  position as a function of the number of tracks associated with the
  primary vertex.  The resolutions obtained with our modeling are
  shown in purple lines while those provided by the ATLAS
  collaboration \cite{ATL-PHYS-PUB-2015-026} are shown in green
  dot-dashed (derived from data) and black dotted (derived from MC
  samples) ones.  }
  \label{fig:resolution}
\end{figure}

A similar analysis, {\it i.e.}, track-based reconstruction of primary vertex
in proton-proton collision, 
has been already performed by the ATLAS \cite{ATLAS:2010lca,
  *Aaboud:2016rmg, ATL-PHYS-PUB-2015-026} and CMS
\cite{Chatrchyan:2014fea, *CMS-DP-2016-041} collaborations
from which we can estimate the accuracy of the
determination of the vertex position at the LHC.  In
Ref.~\cite{ATL-PHYS-PUB-2015-026}, charged tracks with $p_{\rm T}>400\
{\rm MeV}$ were used to reconstruct the primary vertex.  In
Fig.~\ref{fig:resolution}, we show the vertex resolutions to $x$- and
$z$-directions (corresponding to the directions perpendicular and
parallel to the beam axis) provided by the ATLAS collaboration
\cite{ATL-PHYS-PUB-2015-026}; the green (dot-dashed) and black
(dotted) lines show the data and Monte Carlo (MC) results.  Thus, if a
sizable number of charged tracks are associated with the vertex, we
expect that the vertex position is reconstructed with the accuracy of
${\cal O}(10)\ \mu {\rm m}$.  This fact indicates that, if the
distance between two decay vertices is longer than ${\cal O}(10)\ \mu
{\rm m}$ in the pair production process of metastable particles, it
may be possible to distinguish two vertices.  Existence of two
distinct DVs can be used to reduce SM backgrounds, as we discuss
below.

In the following, we quantitatively study how well we can improve the
discovery reach for the new particles with the reconstruction of DVs.
For this purpose, in our MC analysis, we implement an algorithm to
reconstruct DVs using charged tracks.  As we mentioned, we mainly
focus on DVs which are away from the interaction point by $\lesssim
1$~mm, though our method can be used for more displaced cases as well.
Our strategy to reconstruct DVs relies on tracking performance of
charged-particle tracks in the inner detector.

The tracking performance of the ATLAS inner detector for $\sqrt{s} =
13$~TeV is given in Ref.~\cite{ATLAS:trk,
  *ATL-PHYS-PUB-2015-018}.\footnote
{Before the LHC Run-II started, the insertable B-layer (IBL)
  \cite{Capeans:1291633} was installed, which improves the performance
  of track and vertex reconstruction. }
To reproduce the performance of track reconstruction, we shift each track
in parallel by impact parameters.  We neglect the effect
of the curvature of the tracks in this procedure since we focus on DVs
which are very close to the interaction point. We also neglect the
track parameter resolutions regarding its direction, {\it i.e.}, the
azimuthal angle $\phi$ and the polar angle $\theta$, as their
resolutions are sufficiently small: $\sigma_{\phi} \sim 100\,{\rm \mu
  rad}$ and $\sigma_{\cot \theta} \sim 10^{-3}$ \cite{Aad:2009wy}.
Thus, we only consider the resolutions of the transverse and
longitudinal impact parameters, $d_0$ and $z_0 \sin\theta$,
respectively.  Effects of these are taken into account by random
parallel shift of each track.  The resolutions of the impact
parameters depend on the transverse momentum $p_{\rm T}$ and the
pseudorapidity $\eta$ of the track.  In the processes we consider in
this letter, jets have relatively high $p_{\rm T}$ and do not have any
preference for the small polar angle regions. In addition, it is found
that the $\eta$ dependences of the resolutions of the impact
parameters become sufficiently small for $p_{\rm T} \gtrsim$ a few GeV
\cite{Aad:2009wy, impres2015}. For these reasons, we neglect the
$\eta$ dependences of the resolutions in this analysis.  Following
Ref.~\cite{Aad:2009wy}, we parametrize the $p_{\rm T}$ dependence of
the track impact parameter resolutions as
\begin{align}
  \sigma_X \left( p_{\rm T} \right) = \sigma_X \left( \infty \right)
  \left( 1 \oplus p_X / p_{\rm T} \right)~,
\label{eq:resol}
\end{align}
(with $X=d_0$ and $z_0\sin\theta$) where $\sigma_X ( \infty )$ and
$p_X$ are parameters. We determine the values of $\sigma_X(\infty)$
and $p_X$ by fitting this expression onto the $p_{\rm T}$ dependence
of the track impact parameter resolutions measured by the ATLAS
collaboration \cite{ATLAS:trk}.

Next, let us describe the procedure of the vertex reconstruction used
in our analysis, which gives the best-fit point of the vertex for a
given set of charged tracks.  We follow the prescription given in
Refs.~\cite{ATLAS:2010lca, *Aaboud:2016rmg}. In this prescription, the
adaptive vertex fitting algorithm \cite{Fruhwirth:2007hz}, which we
briefly review in Appendix, is exploited to determine the vertex
position.  At the outset of this algorithm, for a given set of charged
tracks, a vertex seed is found from the crossing points of the
reconstructed tracks by means of a method called the fraction of
sample mode with weights (FSMW) \cite{Bickel20063500}. Once the
initial vertex is fixed, we assign a weight, which is given in
Eq.~\eqref{eq:wdef}, to each track such that tracks far from the
vertex point are down-weighted. We then determine another vertex
position at which an objective function, which corresponds to the
vertex $\chi^2$ multiplied by the above weights, is minimized. We
iterate this $\chi^2$ fitting steps with varying a parameter for the
weight assignment until the vertex position converges within
1~$\mu$m. The parameters in this algorithm are set to be the default
values given in Ref.~\cite{Fruhwirth:2007hz} and references therein,
though the results are rather insensitive to these parameters.

To validate our modeling of impact-parameter resolutions and the
vertex reconstruction, we reconstruct the position of primary vertices
in minimum-bias events using our procedure. We generate 47,000
minimum-bias event MC samples with {\tt PYTHIA\,v8.2} \cite{Sjostrand:2007gs}. Here, we use
only tracks with $p_{\rm T} > 400$~MeV and $|\eta|<2.5$ in accordance
with the ATLAS study \cite{ATL-PHYS-PUB-2015-026}.  (For this choice
of minimal $p_{\rm T}$, the best-fit values of $\sigma_{d_0}(\infty)$
($\sigma_{z_0\sin\theta}(\infty)$) and $p_{d_0}$ ($p_{z_0\sin\theta}$)
in Eq.~\eqref{eq:resol} are $30~{\rm \mu m}$ ($90~{\rm \mu m}$) and
$2.1$~GeV ($1.0$~GeV), respectively.)  We then evaluate the
resolutions of primary vertices as a function of the number of tracks.
The results are also shown in Fig.~\ref{fig:resolution}. As can be
seen from this figure, our result is in good agreement with the ATLAS
results \cite{ATL-PHYS-PUB-2015-026}.  Thus, in the following
analysis, we use the above-mentioned procedure to determine the
best-fit points of the decay vertices of pair-produced new particles.

\section{Extending the Reach with DVs}
\label{sec:gluino}

Now, we discuss how and how well the reach for the new physics can be
extended by using the information about the DVs.  We are interested in
the case where
\begin{itemize}
\item[(a)] the metastable particles are pair produced, and
\item[(b)] the metastable particles decay into SM colored particles
  ({\it i.e.}, quarks and/or gluons) as well as possibly other
  particles.
\end{itemize}

In the pair production processes of new metastable particles, no hard
particles are produced at the interaction point (assuming that the new
particles decay after flying sizable amount of distance) except those
from initial state radiation. For this reason, we do not try to
determine the position of the interaction point in each event.\footnote{
We however note that the reconstruction of the primary vertex is 
possible if hard jets or leptons are associated with the production point.
It may also be possible to reconstruct the primary vertex using initial state radiation.
Information about the primary vertex may also be utilized to eliminate the background.
}
We instead use the distance between the two reconstructed DVs,
$|\bm{r}_{\rm DV1} - \bm{r}_{\rm DV2}|$, as a discriminating variable
in our study, where $\bm{r}_{\rm DV1}$ and $\bm{r}_{\rm DV2}$ are
positions of the reconstructed vertices.  As we demonstrate below, we
may extend the LHC reach for new particles by combining conventional
kinematical cuts with the new cuts based on $|\bm{r}_{\rm DV1} -
\bm{r}_{\rm DV2}|$.

Although the strategy we propose is applicable to the class of models
satisfying the conditions (a) and (b) mentioned above, a quantitative
study needs to be performed on a model-by-model basis.  Thus, we
consider metastable gluino as an example, and discuss the implication
of the study of the sub-millimeter DVs.

\subsection{Gluino properties}

Before discussing the LHC search for the metastable gluino with DVs,
we summarize gluino properties which are important for the following
discussion.

A gluino decays through the exchange of
squarks. If squarks $\widetilde{q}$ are heavier than gluino
$\widetilde{g}$, and also if a neutralino $\widetilde{\chi}^0$ and/or
a chargino $\widetilde{\chi}^\pm$ have a mass sufficiently smaller
than the gluino mass, then the tree-level three-body decay processes
$\widetilde{g} \to \bar{q}^\prime q \widetilde{\chi}^{0, \pm}$
dominate the two-body one $\widetilde{g} \to \widetilde{\chi}^0 g$,
which occurs at one-loop level.  The decay length of gluino strongly
depends on the masses of the squarks exchanged in the tree-level
three-body decay processes.  Assuming that the first-generation
squarks are sufficiently lighter than the second- and third-generation
ones, the decay length of the gluino is approximately given by
\cite{Toharia:2005gm, *Gambino:2005eh, *Sato:2012xf}
\begin{equation}
  c\tau_{\tilde{g}} \simeq 200~\mu{\rm m}
  \times \biggl(\frac{m_{\tilde{q}}}{10^3~{\rm TeV}}\biggr)^4
  \biggl(\frac{2~{\rm TeV}}{m_{\tilde{g}}}\biggr)^5 ~,
  \label{eq:ctaug}
\end{equation}
where $m_{\tilde{g}}$ is the gluino mass, $m_{\tilde{q}}$ is the
masses of all the first-generation squarks (which are assume to be
degenerate).  In addition, the masses of bino and wino are assumed to
be much smaller than the gluino mass, while the higgsino mass is
assumed to be larger than the gluino mass.  Note that the above
expression should be multiplied by a factor of $\simeq 1/3$ if squarks
in all generations are degenerate in mass.  Eq.\ \eqref{eq:ctaug}
indicates that the gluino decay length can be as long as $\gtrsim
100~\mu{\rm m}$ for the PeV-scale squarks.  Such heavy squarks,
especially heavy stops, are in fact motivated by the measured value of
the mass of the SM-like Higgs boson, $m_h \simeq 125$~GeV
\cite{Aad:2015zhl}, as a large radiative correction from heavy stops
can easily raise the Higgs-boson mass from its tree-level value
\cite{Okada:1990vk,*Okada:1990gg,*Ellis:1990nz, *Haber:1990aw,
  *Ellis:1991zd}, which is predicted to be smaller than the $Z$-boson
mass \cite{Inoue:1982ej, *Flores:1982pr} in the minimal supersymmetric
SM (MSSM).

Even though the squark masses are at the PeV scale, gluino can still
be around the TeV scale in a technically natural way since the gluino
mass is protected by chiral symmetry. We may find a simple scenario
for the mediation of SUSY-breaking to assure such a split mass
spectrum \cite{Wells:2003tf, *ArkaniHamed:2004fb, *Giudice:2004tc,
  *ArkaniHamed:2004yi, *Wells:2004di, Hall:2011jd, *Hall:2012zp,
  *Ibe:2011aa, *Ibe:2012hu, *Arvanitaki:2012ps, *ArkaniHamed:2012gw,
  *Evans:2013lpa, *Evans:2013dza}.  For example, if all of the
SUSY-breaking fields in the hidden sector are charged under some
symmetries, then the dimension-five operators that give rise to the
gaugino masses are forbidden. In this case, the gaugino masses are
mainly induced by quantum effects, such as the anomaly mediation
effects \cite{Randall:1998uk, Giudice:1998xp} and threshold
corrections at the SUSY-breaking scale \cite{Giudice:1998xp,
  Pierce:1996zz}, and are suppressed by a loop factor compared with
the gravitino mass $m_{3/2}$.  The squark masses are, on the other
hand, generated by dimension-six K\"{a}hler-type operators. If these
operators are induced at the Planck scale, the squark masses are
expected to be ${\cal O}(m_{3/2})$, while if they are induced at a
lower scale, then the resultant squark masses become heavier.
Motivated by this consideration, we regard the squark masses, and thus
$c\tau_{\tilde{g}}$ as well through Eq.~\eqref{eq:ctaug}, as free
parameters in the following analysis.

\subsection{Gluino search with DVs}
\label{sec:gluinoDVsearch}

Now we discuss the gluino search with DVs.  In this letter, in order
to demonstrate that the LHC reach for gluino can be extended with the
information about DVs, we impose a DV-based cut on top of the event
selection criteria used in the ordinary gluino search.  Because
DV-based cut may significantly reduce the SM backgrounds, one had
better optimize the cut parameters to maximize the reach for new
physics.  Such an issue is beyond the scope of this letter, and we
leave it for future study \cite{IJMNH}.

In gluino searches, we focus on events with relatively high-$p_{\rm
  T}$ jets.  In reconstructing DVs, this allows us to tighten the
track selection cuts to $p_{\rm T} > 1$~GeV in order to eliminate
low-$p_{\rm T}$ tracks, whose impact-parameter resolutions are rather
poor as can be seen from Eq.~\eqref{eq:resol}.  (For tracks with
$p_{\rm T} > 1$~GeV, we found that the best-fit values of
$\sigma_{d_0}(\infty)$ ($\sigma_{z_0\sin\theta}(\infty)$) and
$p_{d_0}$ ($p_{z_0\sin\theta}$) in Eq.~\eqref{eq:resol} are $23~{\rm
  \mu m}$ ($78~{\rm \mu m}$) and $3.1$~GeV ($1.6$~GeV), respectively.)
We also require the tracks used for DV reconstruction to satisfy
$|d_0|<10$ mm and $|z_0|<320$ mm \cite{Aad:2015uaa}.

For DV reconstruction, we only use tracks associated with four-highest
$p_{\rm T}$ jets.\footnote
{This reflects the event topology under consideration;
gluinos are always pair-produced and each of them decays into two quarks
and a neutralino.}
If one of these jets contains no track satisfying the above
requirements, then we add the fifth-highest $p_{\rm T}$ jet. If more
than one jets among these five high $p_{\rm T}$ jets do not offer any
tracks which meet the above conditions, then we suppose that DV
reconstruction is not possible in such an event.  Since we do not know
which pair of jets originate from a common parent gluino, we study all
possible patterns of pairings out of the four jets. For each paring,
we find two DVs, each of which is reconstructed from tracks associated
with the corresponding jet pair. Among the possible pairings, we adopt
the one which minimizes an objective function that is defined by the
sum of the weighted vertex $\chi^2$ divided by the sum of the weights
over the two DVs, where we use the same weight as that given in
Ref.~\cite{Fruhwirth:2007hz} (see Eq.~\eqref{eq:chi2def} in Appendix
for a concrete expression). We regard the vertices reconstructed for
this jet pairing as the reconstructed DVs in the following analysis.

In order to see how the variable $|\bm{r}_{\rm DV1} - \bm{r}_{\rm
  DV2}|$ distributes, we perform MC simulation for the gluino pair
production processes.  We first fix the mass and the decay length
$c\tau_{\tilde{g}}$ of gluino (as well as other MSSM parameters).
Then, event samples for the gluino pair production process are
generated; {\tt MadGraph5\_aMC@NLO\,v2}~\cite{Alwall:2014hca} and {\tt
  PYTHIA\,v8.2} 
   are used for this purpose.  We
generate 50,000 events for each mass and lifetime of gluino.  For each
event, we determine the flight lengths of two gluinos using the
lifetime of the gluino, and hence two decay vertices.  The production
point of  each final-state particle is shifted by the flight length of its parent
gluino. Signal event samples are normalized according to the NLL$+$NLO
gluino pair production cross section \cite{Borschensky:2014cia}.
The produced gluinos are forced to
decay into first-generation quarks and a neutralino with a mass of
100~GeV; we refer to these samples as ``light flavor samples.''  After
a fast detector simulation with {\tt DELPHES\,v3}
\cite{deFavereau:2013fsa}, we select only event samples in the signal
region of {\tt Meff-4j-2600} defined in the ATLAS gluino search
\cite{ATLAS:2016kts}, which adopts events with $E_{\rm T}^{\rm
  (miss)}>250\ {\rm GeV}$ (with $E_{\rm T}^{\rm (miss)}$ being the
missing energy), $p_{\rm T} (j_1)>200\ {\rm GeV}$ (with $p_{\rm T}
(j_i)$ being the transverse momentum of $i$-th jet), $p_{\rm T}
(j_4)>150\ {\rm GeV}$, $\Delta\phi (j_{1,2,3,4},E_{\rm T}^{\rm
  (miss)})_{\rm min}>0.4$ (with $\Delta\phi$ being the azimuthal angle
between the jet and the missing energy), aplanarity larger than
$0.04$, $E_{\rm T}^{\rm (miss)}/m_{\rm eff}(4)>0.2$ (with
$m_{\rm eff}(4)$ being the scalar sum of $E_{\rm T}^{\rm (miss)}$ and
the transverse momenta of leading $4$-jets), and $m_{\rm eff}({\rm
  incl.})>2600\ {\rm GeV}$ (with $m_{\rm eff}({\rm incl.})$ being the
scalar sum of $E_{\rm T}^{\rm (miss)}$ and the transverse momenta of
jets with $p_{\rm T}>50\ {\rm GeV}$).

In order to discuss the discovery reach, we should also consider
backgrounds.  As we have mentioned, we require the presence of DVs on
top of the event selection conditions used in the ordinary gluino
searches. The latter conditions drop most of the SM background, and
thus most of the fake DV events are also expected to be eliminated.
Since we impose relatively tight kinematical selection cuts ({\it
  i.e.}, {\tt Meff-4j-2600}), we expect the properties of the SM
background events relevant to tracking and DV reconstruction, such as
the multiplicity and $p_{\rm T}$ of charged tracks, to resemble those
of the signal events after applying the kinematical selection
cuts. With this expectation, we approximate the background event
samples which pass the kinematical selection cuts by the signal event
samples with $c\tau_{\tilde{g}} = 0$. However, one possible difference
between these two is that the SM background may contain $b$ quarks
while our signal event samples called ``light flavor samples'' only include
the first-generation quarks. $b$ quarks tend to be long-lived and thus
may contribute to background considerably. To take into account this
possibility, we generate event samples called ``heavy flavor
samples,'' where the produced gluinos are forced to decay into $b$
quarks, and use them as background samples to be conservative. We
normalize the cross section of the background events to be $0.20$~fb
as observed in the ATLAS gluino search \cite{ATLAS:2016kts}.  In
addition, since we mainly consider DVs inside the beam pipe, we
neglect background vertices from hadronic interactions in the detector
materials and only consider background vertices which are
mis-reconstructed as displaced ones due to the resolution of track
impact parameters.  With this simplification, we reject an event with
a DV whose reconstructed position is inside the detector materials:
{\it i.e.}, $22~{\rm mm}\le |(\bm{r}_{\rm DV})_{\rm T}| \le 25$~mm,
$29~{\rm mm}\le |(\bm{r}_{\rm DV})_{\rm T}| \le 38$~mm, $46~{\rm mm} \le
|(\bm{r}_{\rm DV})_{\rm T}| \le 73$~mm, $84~{\rm mm} \le |(\bm{r}_{\rm
  DV})_{\rm T}| \le 111$~mm, or $|( \bm{r}_{\rm DV})_{\rm T}|\ge 120~{\rm
  mm}$ \cite{PERF-2007-01, Capeans:1291633, 1748-0221-9-02-C02018,
  1748-0221-11-11-P11020}.

\begin{figure}[t]
  \centering
\subcaptionbox{\label{fig:dv_dist} $|\bm{r}_{\rm DV1} - \bm{r}_{\rm DV2}|$
 distribution}{
  \includegraphics[width=0.49\columnwidth]{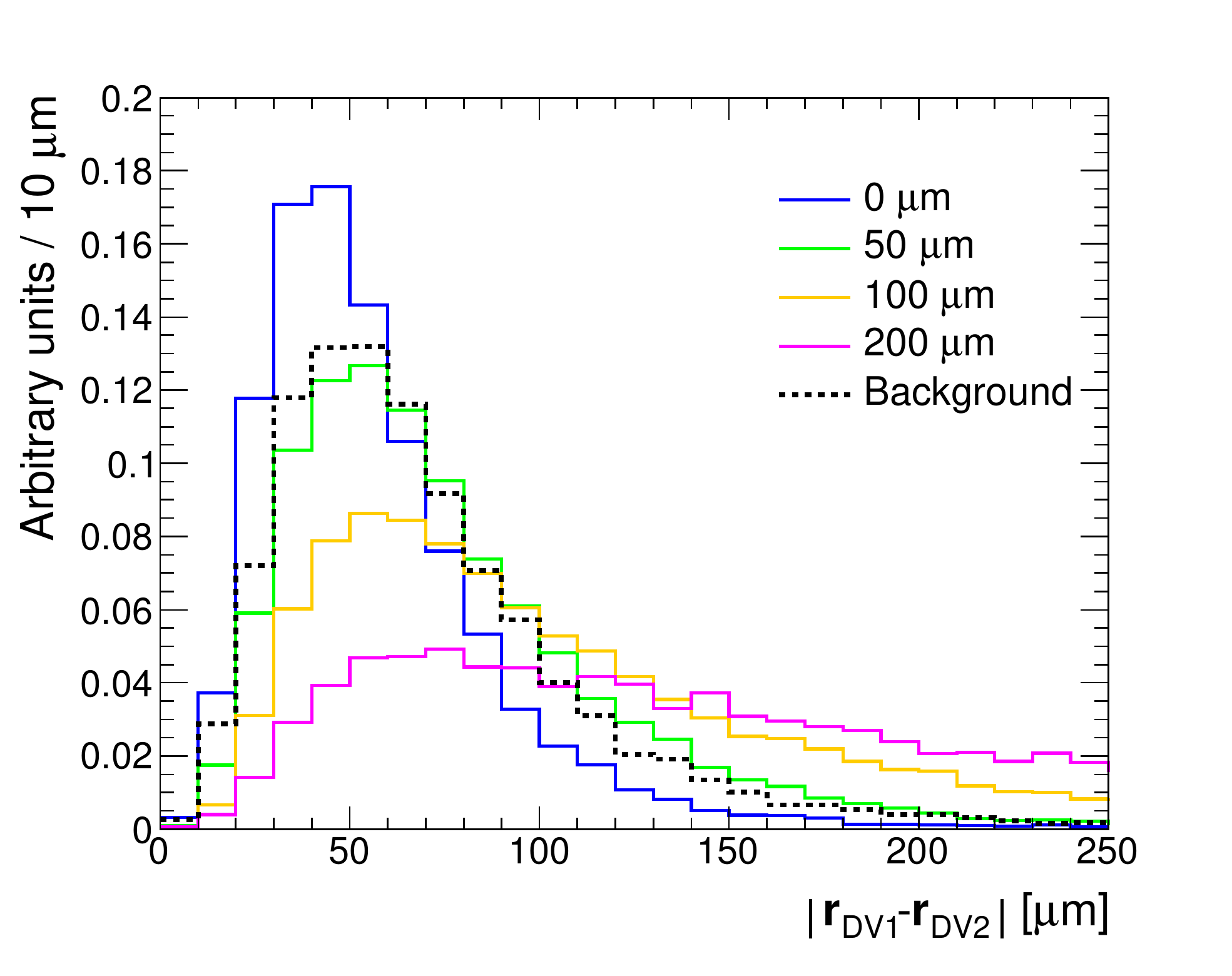}}
\subcaptionbox{\label{fig:dv_eff}
 Efficiency}{\includegraphics[width=0.49\columnwidth]{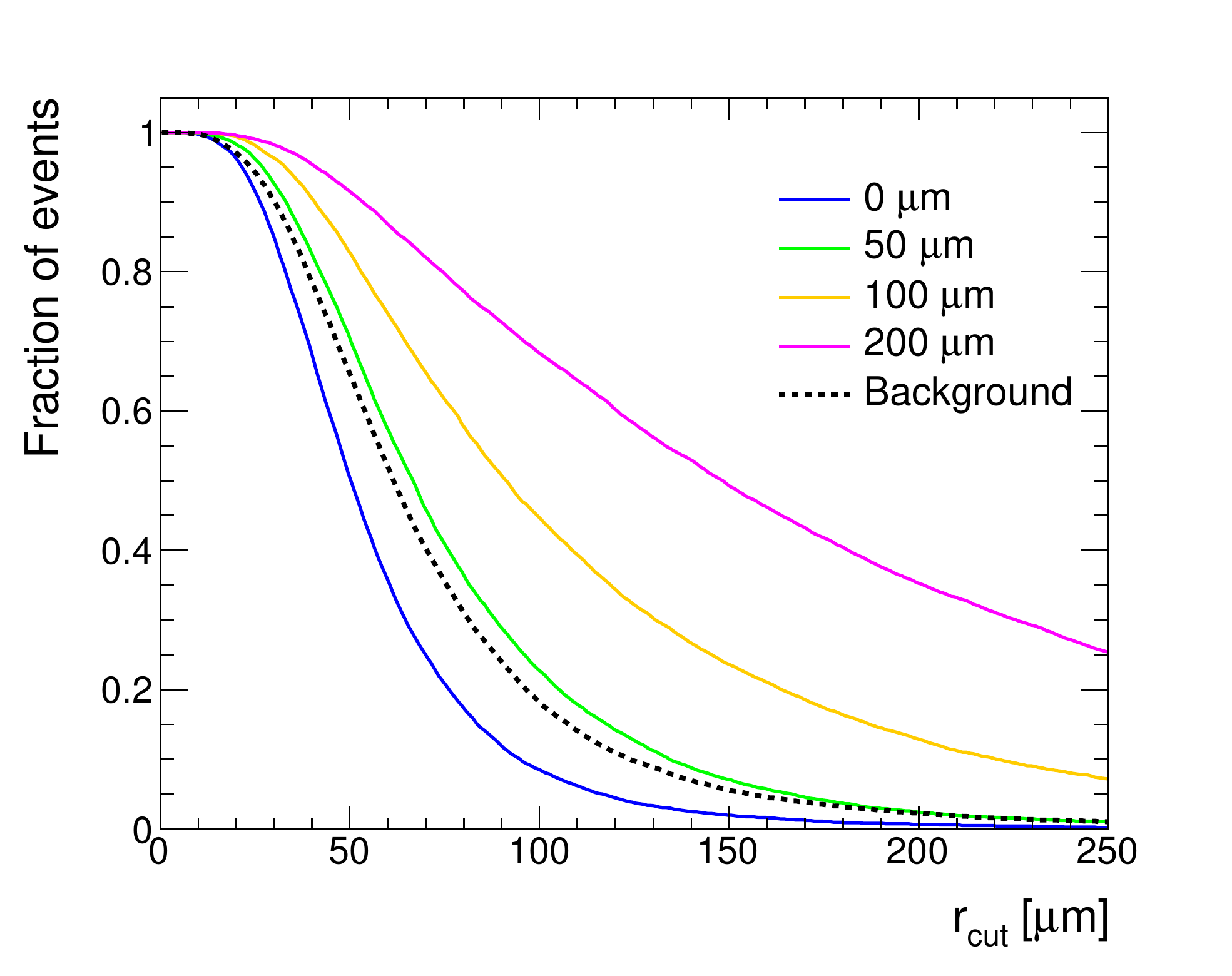}} 
  \caption{\small (a) $|\bm{r}_{\rm DV1} - \bm{r}_{\rm DV2}|$ distribution
in the signal region {\tt Meff-4j-2600} for various $c\tau_{\tilde
 g}$. (b) the fraction of events passing  the selection cut of
 $|\bm{r}_{\rm DV1} - \bm{r}_{\rm DV2}| > r_{\rm cut}$ as a function of
 $r_{\rm cut}$. 
 Here we set $m_{\tilde{g}} = 2.2$~TeV in both figures.
   }
\end{figure}


In Fig.~\ref{fig:dv_dist}, we show the $|\bm{r}_{\rm DV1} -
\bm{r}_{\rm DV2}|$ distribution in the signal region {\tt
  Meff-4j-2600}.
 In addition, in Fig.~\ref{fig:dv_eff} we plot the fraction of events
passing the selection cut of $|\bm{r}_{\rm DV1} - \bm{r}_{\rm DV2}| >
r_{\rm cut}$ as a function of $r_{\rm cut}$. 
Note that the background distribution deviates from the signal distribution with 
 $c\tau_{\tilde g} = 0$ because the background contains $b$ hadrons in jets.
These figures show that
if we set $r_{\rm cut}$ to be $\gtrsim 100~\mu{\rm m}$, then a
significant fraction of SM background fails to pass the selection cut
while a sizable number of signal events for $c\tau_{\tilde{g}} \gtrsim
100~\mu{\rm m}$ remain after the cut. This observation indicates that
this cut may be useful to probe a gluino with a decay length of
$c\tau_{\tilde{g}} \gtrsim 100~\mu{\rm m}$.

To demonstrate the performance of the new selection cut based on DVs,
we show how far we can extend the discovery reach and exclusion limit
of the gluino searches.  We apply the cut to both signal and
background events in the signal region {\tt Meff-4j-2600} and estimate
the expected exclusion and discovery reaches for gluino.  We vary
the cut parameter $r_{\rm cut}$ from $0 \,{\rm \mu m}$ to $1000 \,{\rm
  \mu m}$ by $20 \,{\rm \mu m}$, and determine the highest value of
the gluino mass as a gluino mass reach for each $c\tau_{\tilde{g}}$.
For the integrated luminosity of ${\cal L}=100~{\rm fb}^{-1}$
(1000~fb$^{-1}$) at the 13~TeV LHC, we find that $r_{\rm cut} \sim
200~{\rm \mu m}$ ($400~{\rm \mu m}$) yields the best discovery and
exclusion performance for a gluino with $c\tau_{\tilde{g}} \gtrsim
200~{\rm \mu m}$.

\begin{figure}
  \centering
  \includegraphics[width=0.65\columnwidth]{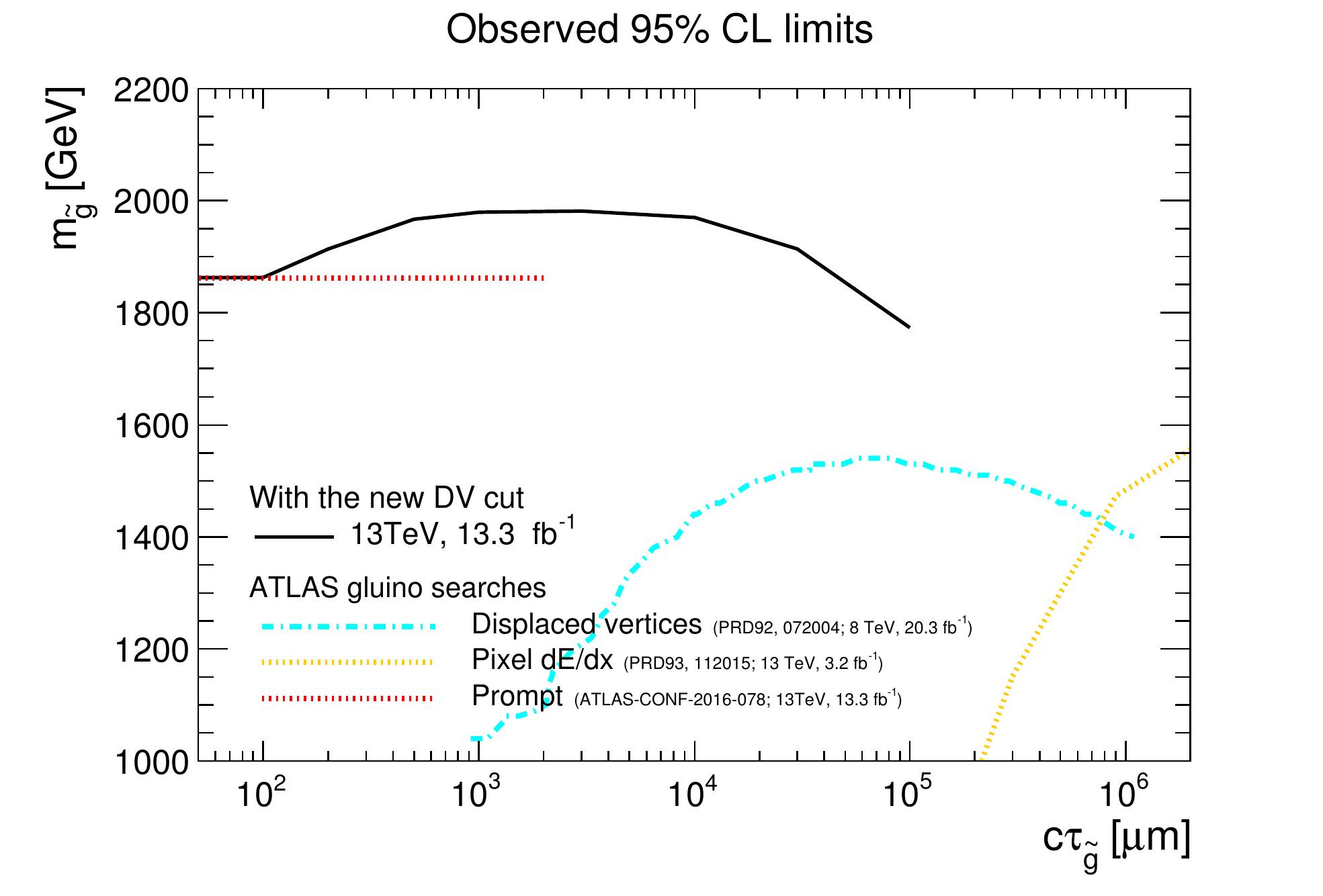}
  \caption{\small The 95\% CL expected exclusion limits on the gluino
    mass with ${\cal L}=13.3\ {\rm fb}^{-1}$ at the 13~TeV LHC run as
    a function of $c\tau_{\tilde g}$ (black solid line). For
    comparison, we also show the 95\% CL exclusion limits given by the
    ATLAS prompt-decay gluino search (red dotted line)
    \cite{ATLAS:2016kts}, the ATLAS DV search (blue dot-dashed line)
    \cite{Aad:2015rba}, and the ATLAS search of large ionization
    energy loss in the Pixel detector (orange dotted line)
    \cite{Aaboud:2016dgf}. }
  \label{fig:exclusionNOW}
\end{figure}

For exclusion limits, we compute the expected 95\% confidence level
(CL) limits on the gluino mass using the $CL_{s}$ prescription
\cite{Read:2002hq, *Junk:1999kv}.  In Fig.~\ref{fig:exclusionNOW}, we
show the expected limit on the gluino mass as a function of
$c\tau_{\tilde g}$ based on the currently available luminosity of
$13.3\ {\rm fb}^{-1}$ at the 13~TeV LHC. We can see that, even with
the current data, the exclusion limit can be improved by about $80$
and $120\ {\rm GeV}$ for $c\tau_{\tilde{g}}=0.3$ and $1\ {\rm mm}$,
respectively.  To compare the result with the current sensitivities of
other gluino searches, we also show the 95\% CL exclusion limits given
by the ATLAS prompt-decay gluino search with the 13~TeV 13.3~fb$^{-1}$
data (red dotted line) \cite{ATLAS:2016kts}, the ATLAS DV search with
the 8~TeV 20.3~fb$^{-1}$ data (blue dot-dashed line)
\cite{Aad:2015rba}, and the ATLAS search of large ionization energy
loss in the Pixel detector with the 13~TeV 3.2~fb$^{-1}$ data (orange
dotted line) \cite{Aaboud:2016dgf}.  The existing metastable gluino
searches are insensitive to a gluino with $c\tau_{\tilde{g}} \lesssim
1$~mm, as shown in Fig.~\ref{fig:exclusionNOW} (blue dot-dashed and
orange dotted lines), to which searches with the new DV cut may offer
a good sensitivity.  In this sense, this new search strategy plays a
complementary role in probing metastable gluinos.

\begin{figure}
  \centering
  \includegraphics[width=0.65\columnwidth]{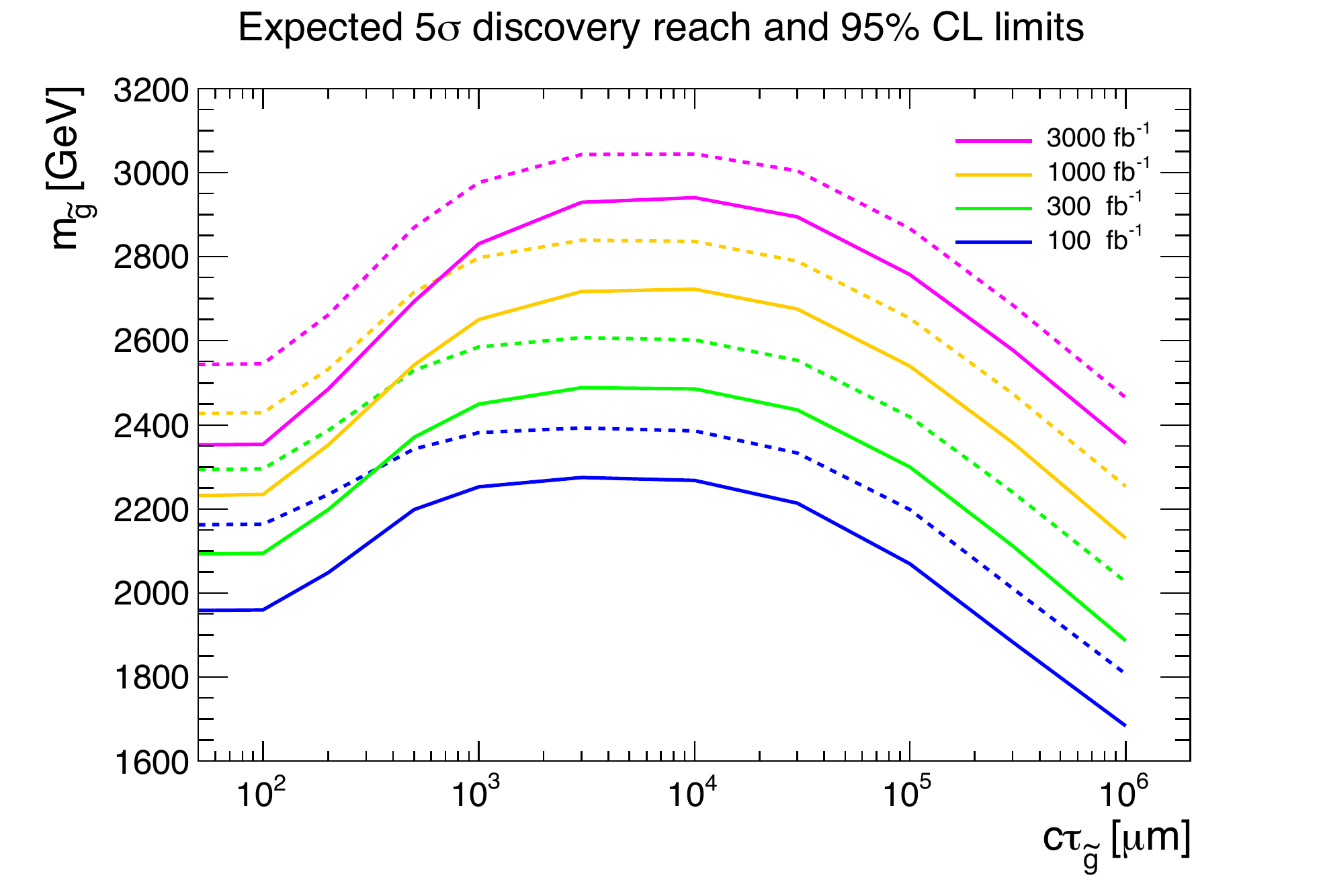}
  \caption{\small The expected 95\% CL exclusion limits (dotted) and
    $5\sigma$ discovery reaches (solid) as functions of $c\tau_{\tilde
      g}$ for different values of integrated luminosity at the 13~TeV
    LHC run.  }
  \label{fig:future}
\end{figure}

To see the future prospect, we also derive the expected 95\% CL
exclusion limits as well as $5\sigma$ discovery reach with larger
luminosity.  The expected discovery reach is determined by calculating
the expected significance of discovery $Z_0$ \cite{Cowan:2010js}:
\begin{align}
Z_0 = \sqrt{2\left\{ \left(S + B\right) \log \left( 1 + S/B \right) -S
 \right\} }~,  
\end{align}
where $S$ ($B$) is the expected number of signal (background) events.
We then require $Z_0$ to be larger than $5$ for discovery.  In Fig.\
\ref{fig:future}, we show the expected 95\% CL exclusion limits and
$5\sigma$ discovery reaches for gluino as functions of $c\tau_{\tilde
  g}$ for different values of integrated luminosity at the 13~TeV LHC
run.  Notice that the expected reaches for an extremely small
$c\tau_{\tilde{g}}$ should correspond to those for the prompt-decay
gluino with the same data set since the new DV cut plays no role in
this case.  As can be seen from this figure, the reach for the gluino
can be extended with the help of the additional DV selection cut for
$c\tau_{\tilde g} \gtrsim 100~\mu{\rm m}$; for instance, for a gluino
with $c\tau_{\tilde g} \sim \mathcal{O}$(1--10)~mm, the expected
discovery reach for the gluino mass can be extended by as large as
$\sim 300$~GeV ($500$~GeV) with an integrated luminosity of ${\cal
  L}=100~{\rm fb}^{-1}$ (1000~fb$^{-1}$).  Because charged tracks with
$|d_0|>10$ mm are not included in the analysis, and also because we
reject all events with a DV whose reconstructed position radius is
larger than $120$~mm, the expected exclusion limits decrease for
$c\tau_{\tilde{g}} \gtrsim 100$~mm. Such a larger $c\tau_{\tilde{g}}$
region can however be covered by other long-lived gluino searches.
(Remember that these numbers are based on the events in the signal
region {\tt Meff-4j-2600}.  For more accurate estimation of the
improvement, one should carefully optimize the selection criteria,
with which we may have better reach.)

\subsection{Lifetime measurement}

If a new metastable particle is discovered at the LHC, measurement of
its lifetime is of crucial importance to understand the nature of new
physics behind this metastable particle. In this subsection, we
discuss the prospect of the lifetime measurement by means of the DV
reconstruction method we have discussed.

\begin{figure}
  \centering
 \subcaptionbox{\label{fig:ctau_sig_0mu} $c\tau_{\tilde{g}}^{\rm (hypo)} = 0~\mu {\rm m}$}{
  \includegraphics[width=0.48\columnwidth]{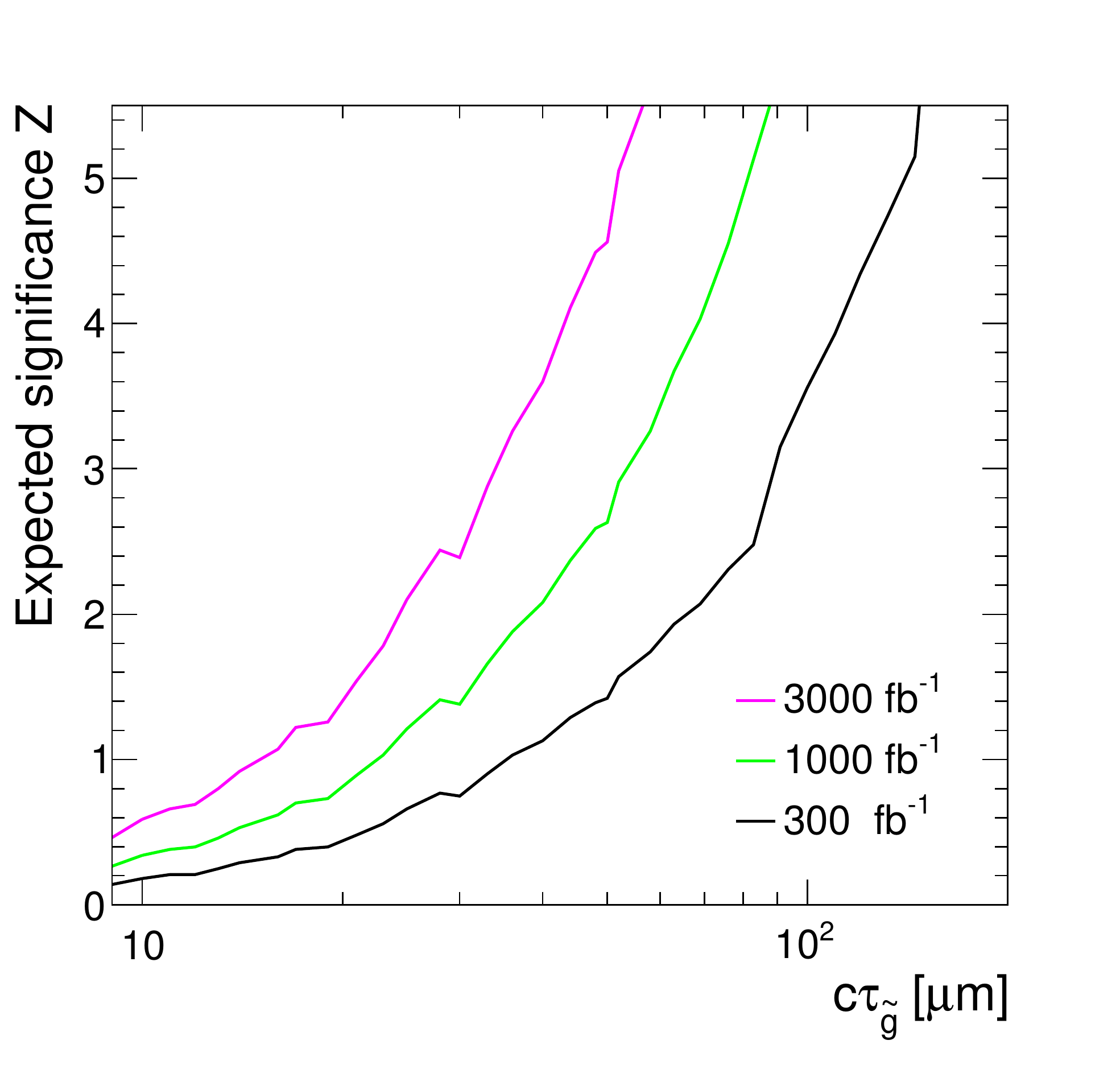}}
\subcaptionbox{\label{ctau_sig_200mu} $c\tau_{\tilde{g}}^{\rm (hypo)} = 200~\mu {\rm m}$}{
  \includegraphics[width=0.48\columnwidth]{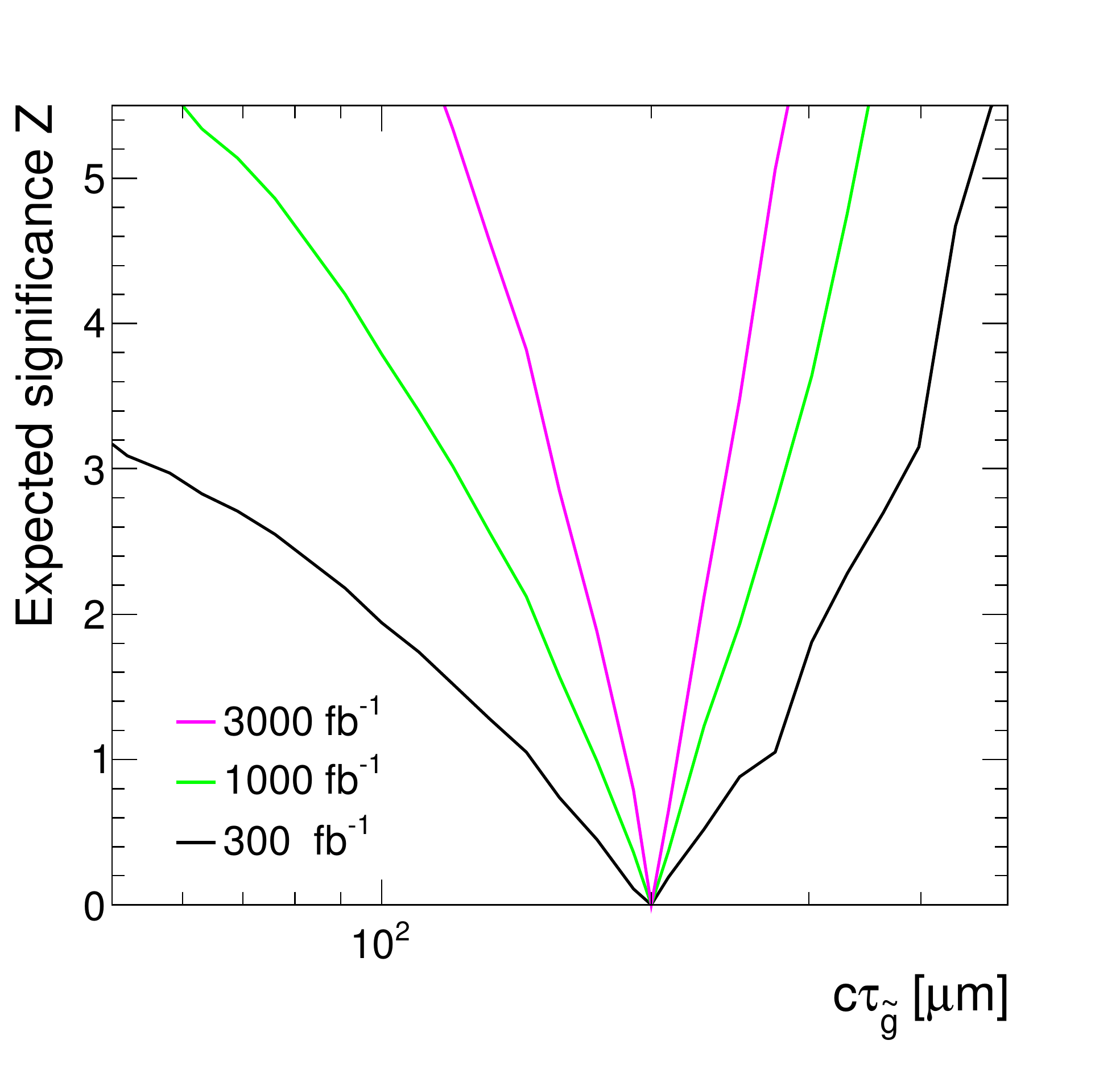}}
  \caption{\small The expected significance of rejection 
  $\langle Z_{c\tau_{\tilde{g}}^{\rm (hypo)}} \rangle_{c\tau_{\tilde{g}}}$
  as a function of 
  $c\tau_{\tilde{g}}$ for an integrated luminosity of 
  $300$,~$1000$, and $3000~{\rm
 fb}^{-1}$ in the black, green, and purple lines, respectively. 
 Here, we set $m_{\tilde{g}} =
 2.2$~TeV. See text for the definition of $\langle
Z_{c\tau_{\tilde{g}}^{\rm (hypo)}} \rangle_{c\tau_{\tilde{g}}}$. 
   }
  \label{fig:ctau_sig}
\end{figure}

\begin{figure}
  \centering
  \subcaptionbox{\label{fig:ctau_bound300} ${\cal L}=300~{\rm fb}^{-1}$}{
    \includegraphics[width=0.48\columnwidth]{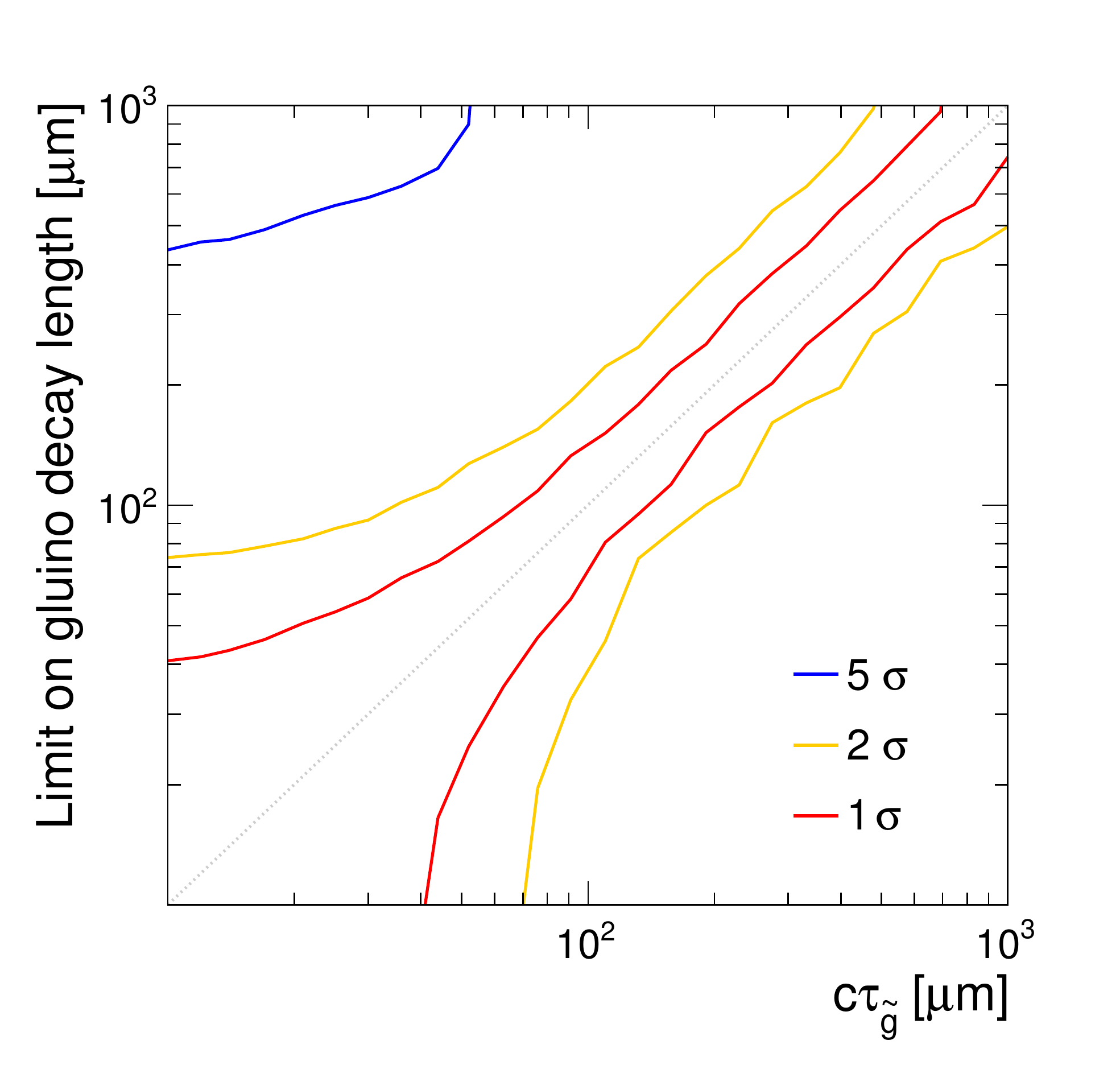}}
  \subcaptionbox{\label{fig:ctau_bound3000} ${\cal L}=3000~{\rm fb}^{-1}$}{
    \includegraphics[width=0.48\columnwidth]{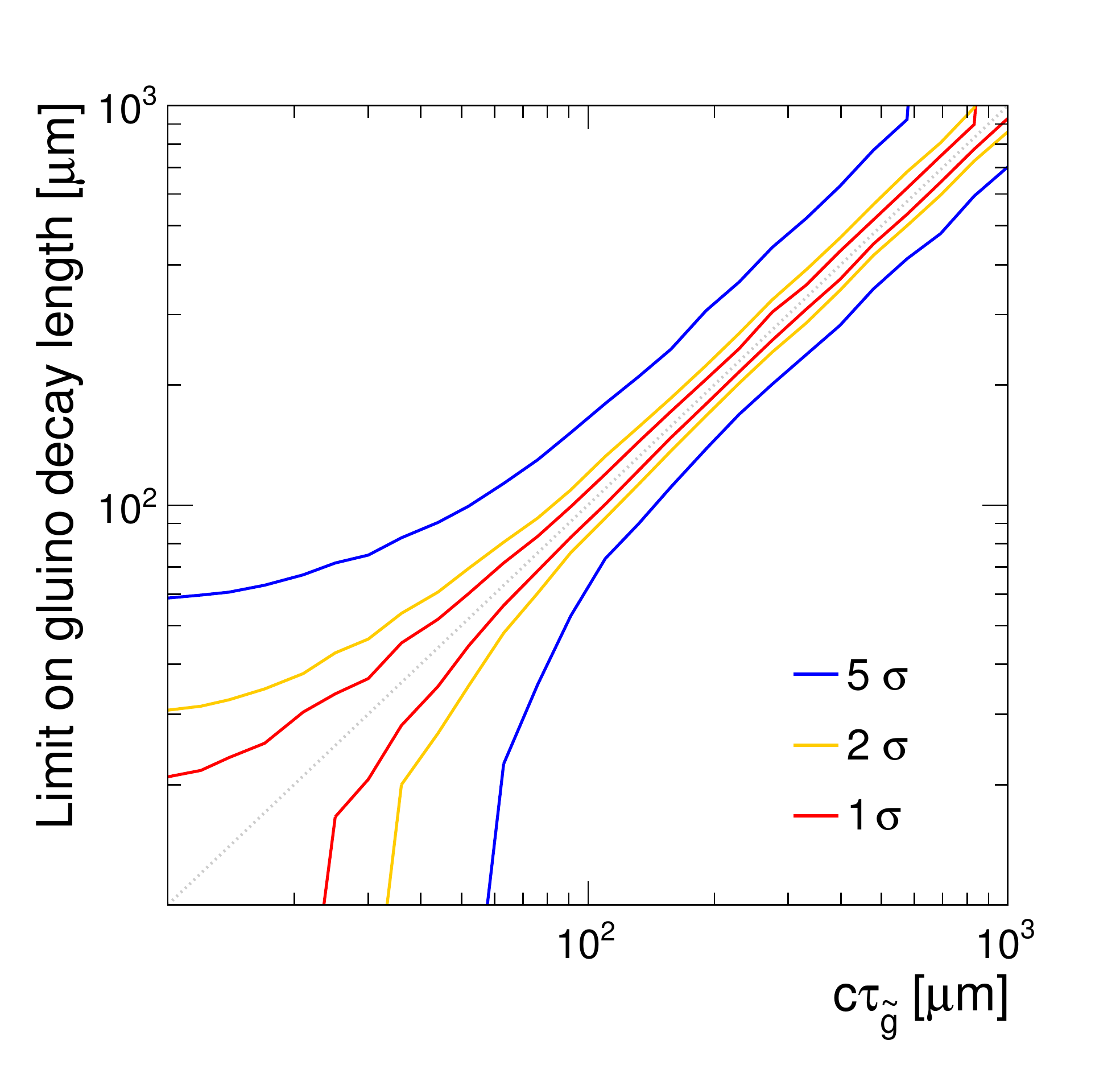}}
  \caption{\small The expected upper and lower bounds decay length of
    gluino as a function of the underlying value of $c\tau_{\tilde
      g}$. Here, we set $m_{\tilde{g}} = 2.2$~TeV.}
\label{fig:ctau_bound}
\end{figure}

To see the prospect of the lifetime measurement, 
we study the expected
significance of rejection of a hypothesis that the gluino decay length
is $c\tau_{\tilde{g}}^{\rm (hypo)}$ for gluino samples with a decay
length of $c\tau_{\tilde{g}}$. 
Event samples are binned according to the DV distance $|\bm{r}_{\rm DV1} -
\bm{r}_{\rm DV2}|$ of the events.
Then the expected significance $\langle
Z_{c\tau_{\tilde{g}}^{\rm (hypo)}} \rangle_{c\tau_{\tilde{g}}}$ is
determined as $\langle Z_{c\tau_{\tilde{g}}^{\rm (hypo)}}
\rangle_{c\tau_{\tilde{g}}}\equiv\sqrt{
\Delta \chi^2 (c\tau_{\tilde{g}}^{\rm (hypo)},c\tau_{\tilde{g}})}$, where
\begin{align}
\Delta \chi^2 (c\tau_{\tilde{g}}^{\rm (hypo)},c\tau_{\tilde{g}})
= 
\sum_{{\rm bin}\,\, i} 
\frac{\left\{ S_i(c\tau_{\tilde{g}}^{\rm (hypo)}) - S_i(c\tau_{\tilde{g}})
 \right\}^2}{S_i(c\tau_{\tilde{g}}^{\rm (hypo)}) +  B_i}\, .
\label{eq:chisq}
\end{align}
Here, $S_i(c\tau)$ is the expected number of signal events in the bin
$i$ on the assumption that gluinos have a decay length of $c\tau$,
while $B_i$ is the number of SM background. 
We show the expected significance
for $c\tau_{\tilde{g}}^{\rm (hypo)} = 0$ and $200~{\rm \mu m}$ as a
function of the gluino decay length $c\tau_{\tilde{g}}$ 
used to generate the
data sample in Figs.~\ref{fig:ctau_sig_0mu} and \ref{ctau_sig_200mu},
respectively, for a gluino with a mass of 2.2~TeV. Here we use the
visible cross-section $\epsilon \sigma$ of $8.8\times10^{-2}~{\rm fb}$
for signal events, which is defined by the product of the production
cross-section $\sigma$ and the fraction of signal events in the signal
region {\tt Meff-4j-2600} estimated from our fast detector simulation,
$\epsilon$.  In Figs.~\ref{fig:ctau_bound300} and
\ref{fig:ctau_bound3000}, we also show the expected upper and lower
bounds on the decay length as a function of 
$c\tau_{\tilde g}$.  From the figures, we find that a metastable
gluino with $c\tau_{\tilde g} \gtrsim 25~(50)~{\rm \mu m}$ can be
distinguished from a promptly decaying one with the significance of
$2\sigma$ ($5\sigma$) with an integrated luminosity of $3000~ {\rm
  fb}^{-1}$. Moreover, Fig.~\ref{ctau_sig_200mu} shows that the decay
length of a gluino with $c\tau_{\tilde{g}} \sim {\cal O}(100)~\mu {\rm
  m}$ can be measured with an ${\cal O}(1)$ accuracy at the
high-luminosity LHC. With such a measurement, we may probe the squark
mass scale $m_{\tilde{q}}$ via Eq.~\eqref{eq:ctaug} even though
squarks are inaccessible at the LHC, which sheds light on the SUSY
mass spectrum as well as the mediation mechanism of SUSY-breaking
effects.

\section{Conclusions and Discussion}
\label{sec:concl}

In this letter, we have discussed a method of reconstructing DVs that
originate from decay of metastable particles on the assumption that
these metastable particles are always pair-produced and their decay
products contain high-$p_{\rm T}$ jets. We especially consider gluinos
in the SUSY models as an example, which tend to be metastable when
squarks have masses much larger than the TeV scale. It is found that
this method can separate out DVs if the gluino decay length is $\gtrsim
100~\mu {\rm m}$. Then, we have seen that an event selection cut based
on this DV reconstruction may be utilized to improve the potential of
the gluino searches for a gluino with $c \tau_{\tilde{g}} \gtrsim
100~\mu {\rm m}$. In particular, if $c\tau_{\tilde g} \sim
\mathcal{O}$(1--10)~mm, then the exclusion and discovery reaches for the
gluino mass can be extended by about $370$~GeV and $500$~GeV,
respectively, with an integrated luminosity of 1000~fb$^{-1}$ at the
13~TeV LHC. 
Furthermore, 
with an integrated luminosity of 3000~fb$^{-1}$, 
it is possible to measure
the gluino decay length with an ${\cal O}(1)$ accuracy for a gluino with
$c\tau_{\tilde{g}}\sim{\cal O}(100)~\mu {\rm m}$ and $m_{\tilde{g}} =
2.2$~TeV, which may allow us to probe the PeV-scale squarks
indirectly. Although we have concentrated on metastable gluinos in SUSY
models, a similar technique may be used to probe DV signatures from
other metastable particles. A more extensive study will be done
elsewhere \cite{IJMNH}. 

\section*{Acknowledgments}

The authors thank T. Yamanaka and S. Adachi for useful discussion.
This work was supported by the Grant-in-Aid for Scientific Research on
Scientific Research C (No.26400239 [TM]), Young Scientists B
(No.15K17653 [HO]), and Innovative Areas (No.16H06489 [OJ],
No.16H06490 [TM]).

\section*{Appendix: Vertexing Method}
\appendix

\renewcommand{\theequation}{A.\arabic{equation}}
\setcounter{equation}{0}

Here, we give a brief review on the vertexing method exploited in our
analysis, as well as a concrete expression for the objective function
used to determine the reconstructed DVs. 

Our vertexing method is based on the adaptive vertex fitting algorithm
\cite{Fruhwirth:2007hz}. In this algorithm, an initial vertex position
is found using the FSMW method \cite{Bickel20063500} for a pair of
jets in question. This method first defines a crossing point for a pair
of the two tracks chosen from each jet as the closest midpoint of these
tracks. We then assign a weight to this crossing point,
\begin{equation}
 w \equiv \left(d + d_{\rm min}\right)^{-\frac{1}{2}} ~,
\end{equation}  
where $d$ is the distance between the two tracks, and we set $d_{\rm min}
= 10~\mu{\rm m}$ following Ref.~\cite{Fruhwirth:2007hz}. This weight
gets larger if the distance between the two tracks associated with the
crossing point is smaller. Next, for a spatial coordinate, say, the
$x$-coordinate, we
consider a distribution of the crossing points and define a weighted
interval for the distribution as the length of the interval divided by
the sum of the weights of the points in the interval. We then find the
smallest weighted interval that covers at least 40\% of all the
points. This process is recursively performed for the obtained smallest
weighted interval until the interval contains only two points, and
eventually the midpoint of the remaining two points is defined as the
$x$-coordinate of the initial vertex position. We perform this procedure
for each spatial direction.

For the vertex position $\bm{v}$ determined above, we define 
\begin{equation}
 \chi_i^2 (\bm{v}) \equiv \frac{d_i^2 (\bm{v})}{\sigma_{d_0}^2 +
  \sigma_{z_0 \sin\theta}^2} ~,
\end{equation}
for each track $i$, where $d_i (\bm{v})$ denotes its distance from the
vertex $\bm{v}$. We further assign a weight $w_i$ to each track that is
defined by
\begin{equation}
 w_i (\chi_i^2) \equiv \frac{\exp\left(-\chi_i^2/2T\right)}
{\exp\left(-\chi_i^2/2T\right) + \exp\left(-\chi_c^2/2T\right)} ~,
\label{eq:wdef}
\end{equation}
where we use $\chi_c = 3$ \cite{Fruhwirth:2007hz} and $T$ is a parameter
that we choose in the following. As can be seen from this expression, if 
a track is far away from the vertex $\bm{v}$, a fairly small weight is
assigned to the track. Then, we determine a new vertex position by solving
\begin{equation}
 \sum_{i} w_i \left(\chi_i^2 (\bm{v})\right) 
 \chi_i(\bm{v}_{\rm new}) 
\frac{\partial \chi_i (\bm{v}_{\rm new})}{\partial \bm{v}} = 0 ~,
\end{equation}
with respect to $\bm{v}_{\rm new}$. This new vertex position
($\bm{v}_{\rm new}$) is then used as an
initial vertex position to repeat this process. We iterate the above
process with varying the parameter $T$ as $T = 256 \to 64 \to 16 \to 4
\to 1 \to 1\to \dots$ until $T = 1$ and the vertex position converges within
$1~\mu{\rm m}$. 

As mentioned in Sec.~\ref{sec:gluinoDVsearch}, the weight $w_i$ defined
in Eq.~\eqref{eq:wdef} is also used to determine the jet pairing for the
reconstruction of DVs out of four jets. Among the three possible
pairings of the four jets, we choose the one which minimizes 
\begin{equation}
 \chi^2 \equiv \frac{\sum_{i\in {\rm trk} (\bm{v}_1)} w_i \left(\chi_i^2
(\bm{v}_1)\right) \chi_i^2(\bm{v}_1) 
+\sum_{j\in {\rm trk} (\bm{v}_2)} w_j \left(\chi_j^2
(\bm{v}_2)\right) \chi_j^2(\bm{v}_2)   }{
\sum_{i\in {\rm trk} (\bm{v}_1)} w_i \left(\chi_i^2
(\bm{v}_1)\right)
+\sum_{j\in {\rm trk} (\bm{v}_2)} w_j \left(\chi_j^2
(\bm{v}_2)\right)}
~,
\label{eq:chi2def}
\end{equation}
where ${\rm trk} (\bm{v}_{1,2})$ denotes the set of tracks associated
with the DV $\bm{v}_{1,2}$ reconstructed for each pair of jets, and we
take $T = 1$ and $\chi_c = 3$ in the weights. We
define the reconstructed DVs by $\bm{r}_{\rm DV1,2} \equiv \bm{v}_{1,2}$
for the jet pairing that minimizes this $\chi^2$, and use them in our
analysis.

{\small 
\bibliographystyle{aps}
\bibliography{ref}
}


\end{document}